# Hole-induced insulator-to-metal transition in $La_{1-x}Sr_xCrO_3$ epitaxial films

K.H.L Zhang[1], Y. Du[2], P. V. Sushko[1], M. E. Bowden[2], V. Shutthanandan[2], S. Sallis[3], L.F.J. Piper[3], S.A. Chambers[1*]

[1]Physical Sciences Division, Fundamental and Computational Sciences Directorate, Pacific Northwest National Laboratory, Richland, WA 99352 U.S.A

[2]Environmental Molecular Sciences Laboratory, Pacific Northwest National Laboratory, Richland, WA 99352 U.S.A

[3]Materials Science & Engineering, Binghamton University, Binghamton, New York 13902, USA

*Corresponding author: sa.chambers@pnnl.gov

**Abstract**

We have investigated the evolution of the electronic properties of $La_{1-x}Sr_xCrO_3$ ($0 \leq x \leq 1$) epitaxial films deposited by molecular beam epitaxy (MBE) using x-ray diffraction, x-ray photoemission spectroscopy, Rutherford backscattering spectrometry, x-ray absorption spectroscopy, electrical transport, and *ab initio* modeling. $LaCrO_3$ is an antiferromagnetic insulator whereas $SrCrO_3$ is a metal. Substituting $Sr^{2+}$ for $La^{3+}$ in $LaCrO_3$ effectively dopes holes into the top of valence band, leading to $Cr^{4+}$ ($3d^2$) local electron configurations. Core-level and valence-band features monotonically shift to lower binding energy with increasing x, indicating downward movement of the Fermi level toward the valence band maximum. The material becomes a *p*-type semiconductor at lower doping levels and an insulator-to-metal transition is observed at $x \geq 0.65$, but only when the films are deposited with in-plane compression via lattice-mismatched heteroepitaxy. Valence band x-ray photoemission spectroscopy reveals diminution of electronic state density at the Cr 3*d* $t_{2g}$-derived top of the valence band while O K-edge x-ray absorption spectroscopy shows the development of a new unoccupied state above the Fermi level as holes are doped into $LaCrO_3$. The evolution of these bands with Sr concentration is accurately captured using density functional theory with a Hubbard *U* correction of 3.0 eV (DFT + *U*). Resistivity data in the semiconducting regime ($x \leq 0.50$) do not fit perfectly well to either a polaron hopping or band conduction model, but are best interpreted in terms of a hybrid model. The activation energies extracted from these fits are well reproduced by DFT + $U$.

## I. Introduction

Perovskite oxides remain a focus of attention over the past two decades following the discoveries of high-temperature superconductivity in cuprates and colossal magnetoresistance in manganites.[1,2] Hole-doping plays a crucial role in the emergence of these fascinating properties.[1,3] Determining how the electronic and magnetic structures evolve with hole-doping is critical to understanding the rich range of observed phenomena. In many cases, the parent compounds La$MO_3$ ($M$ = Ti - Co) are insulators and some are thought to exhibit strong electron correlation effects.[4,5] Substitution of La$^{3+}$ by Sr$^{2+}$ introduces holes at the top of valence band (VB), thereby driving the system to a metallic state, accompanied by either a Pauli paramagnetic ($M$ = Ti and V) or a ferromagnetic ($M$ = Mn and Co) ground state.[5-9]

Compared to other A-site doped, first-row transition-metal complex oxides, La$_{1-x}$Sr$_x$CrO$_3$ (LSCO) is relatively unexplored, particularly at a fundamental level. LSCO and its derivatives exhibit mixed electronic and ionic conductivity and are of interest for electrodes in solid oxide fuel cells,[10,11] chemical sensors, and thermal and photocatalysts.[12] Understanding the intrinsic electronic, optical, and chemical properties of LSCO is thus of vital importance for this material to be most effectively utilized in these applications. From the relatively spares literature available, it appears that LSCO behaves rather differently than analogous LS$MO$ systems. No insulator-to-metal transition has been found even at the highest reported Sr doping level (x = 0.5),[13-18] in contrast to La$_{1-x}$Sr$_x$$MO_3$ ($M$ = Ti, V, Mn and Co), which exhibit such a transition at x ≤ 0.17.[6-8] Previous investigations of LSCO have utilized polycrystalline powder samples. Webb *et al*.[13]



measured conductivities between 77K and 1300K for samples with x up to 0.2. These authors report room-temperature resistivities ranging from 200 Ω-cm at x = 0.0 to ~0.9 Ω-cm at x = 0.2. Likewise, Karim and Alfred [14] made an analogous set of samples with x up to 0.4, measured transport from 300K to 2000K, and reported room-temperature resistivities ranging from ~500 Ω-cm at x = 0.0 to ~36 Ω-cm at x = 0.4. Fitting $\rho(T)$ vs $T$ data over these rather wide temperature ranges using both standard semiconductor ( $\rho \propto \exp(\frac{Ea}{kT})$ ) and polaron hopping ( $\rho \propto T\exp(\frac{Ea}{kT})$ ) models revealed clearly better agreement for polaron hopping. In this scenario, $Cr^{4+}$ cations accompanying Sr doping tend to distort the surrounding lattice, thereby trapping holes in the associated potential wells and becoming somewhat conductive. The activation energy for small polaron hopping was estimated to be 0.1–0.2 eV depending on the Sr concentration.

Spectroscopic investigations of occupied and unoccupied bands of LSCO using highly surface sensitive ultraviolet photoemission spectroscopy (UPS) and bremsstrahlung isochromat (BI) spectroscopy, respectively, showed only very slight movement of the highest (lowest) occupied (unoccupied) band toward the Fermi level in going from x = 0.1 to x = 0.5. Significantly, a gap of ~2.5 eV remained at x = 0.5.[15] Maiti *et al.*[15] also reported that LSCO turns black for very low Sr doping level (x = ~0.1%). A closely-related study involving x-ray photoelectron spectroscopy (XPS) and x-ray absorption spectroscopy (XAS) concluded that the doped holes remain localized on $Cr^{4+}$ sites, in support of the small polaron model.[16] These previous studies raise a number of questions with regard to fundamental properties of LSCO and pertaining to other hole-doped complex oxides.



1. Are the relatively high resistivities and activations energies for electronic transport intrinsic to LSCO, or the result of grain boundaries or other defects in these polycrystalline specimens? (The fact that pure LCO, an insulator in the bulk, was reported to exhibit measurable conductivity at ambient temperature suggests the presence of oxygen vacancies and/or electrically active impurities, both which dope the material.)
2. Are the doped holes in LSCO indeed trapped on $Cr^{4+}$ sites, or are they itinerant? How can polycrystalline LSCO turn black for as little as ~0.1 at. % Sr doping (x = 0.001) and yet remain nonmetallic up to x = 0.5?
3. Are the electrical properties of epitaxial LSCO at lower temperatures (100 – 300K) best described in terms of polaronic hopping, as are those for polycrystalline LSCO at higher temperatures (300 – 2000K)?
4. How important is electron correlation in LSCO?

Our goal in the present work is to answer these and other fundamental questions by the preparing high-purity epitaxial films of $La_{1-x}Sr_xCrO_3$ over the full range of composition ($0 \leq x \leq 1$) and to determine the associated intrinsic electronic properties. Combining LCO and SCO to make epitaxial LSCO and understanding the results is aided by detailed understanding of the end members. One end member, $LaCrO_3$ (LCO), is a G-type antiferromagnetic insulator.[4] A long-standing debate in the literature regarding the properties of LCO was recently resolved by a combined experimental and theoretical investigation which revealed that LCO deposited by molecular beam epitaxy (MBE) exhibits a gap of 2.8 eV arising from intra-Cr *d-d* transitions, and that the O 2*p*-Cr 3*d* charge-transfer gap is between 4.7 and 5.0 eV, depending on strain.[19] The properties of the other end member, $SrCrO_3$ (SCO), have also been controversial. SCO was first claimed to be a paramagnetic metallic oxide,[20] but a more recent pressure-dependent



electrical resistivity study indicated that it is an insulator at ambient pressure and undergoes a transition to a metal with increasing pressure.[21] SCO should be metallic by virtue of the presence of holes at the top of the VB, which is strongly Cr 3$d$ derived, viz. $Cr^{4+}$ $(t_{2g})^2$. By preparing model epitaxial films of SCO by MBE and performing detailed characterization, Zhang et al.[22] revealed the intrinsic properties of this material. Even when deposited with sufficiently high oxygen flux to fully oxidize Cr in evolving LCO films, MBE-grown SCO was shown to be markedly oxygen deficient and nucleate as a mixture of rhombohedral, semiconducting $SrCrO_{2.8}$ (R-SCO) and cubic, perovskite $SrCrO_3$ (P-SCO). The driving force behind R-SCO formation is that $Cr^{4+}$ is more stable in tetrahedral than octahedral coordination, as judged by the metastability of bulk oxides containing $Cr^{4+}$ in octahedral coordination. R-SCO(001) epitaxial films contain {111}–oriented planes of $SrO_2$ with adjacent layers of tetrahedrally coordinated $Cr^{4+}$. These O-deficient planes act as fast diffusion paths for $O^{2-}$ anions, leading to relatively low-temperature oxidation (~250°C) of R-SCO to the perovskite structure (P-SCO) in air. Although the oxidation kinetics for SCO are too slow for complete oxidation in the MBE pressure regime, total oxidation readily occurs at atmospheric pressure at rather low temperatures.

Our epitaxial $La_{1-x}Sr_xCrO_3$ films ($0 \leq x \leq 1$) were deposited on $LaAlO_3(001)$ substrates using MBE with the attendant high degree of control over film structure and composition. In contrast to previous studies on polycrystalline samples going up to x = 0.5 which were found to be highly resistive and yielded no hint of an insulator-to-metal transition, our epitaxial films are considerably more conductive for all nonzero doping levels, approach metallicity at x = 0.50, and clearly undergo an insulator-to-metal transition at x = 0.65. Moreover, our pure LCO is so highly resistive that it exhibits no measureable conductivity, as expected due to its insulating character of bulk material, indicating high purity and negligible oxygen vacancy conditions. Valence band



XPS and O K-edge XAS reveal a clear transfer of spectral weight from the occupied Cr $3d$ $t_{2g}$ valence band to a split-off unoccupied state above the Fermi level of the same symmetry with hole doping. This result is also predicted by the density functional simulations utilizing the general gradient approximation (GGA) and Hubbard $U$ correction.

## II. Experimental and modeling methods

Epitaxial $La_{1-x}Sr_xCrO_3$ films with x = 0, 0.12, 0.25, 0.50, 0.65, 0.75, 0.85 and 1.0 and thicknesses of 20-80 nm were grown on $LaAlO_3$(001) substrates by MBE. The substrates were loaded into an ultrahigh vacuum chamber and cleaned at 700 °C for 20 min in an oxygen partial pressure of $6.0 \times 10^{-6}$ Torr prior to film growth. La, Sr and Cr were evaporated from high-temperature effusion cells and evaporation rates were calibrated each time prior to growth using a quartz crystal oscillator (QCO) positioned at the substrate position. The substrate temperature was set to 700 °C and the $O_2$ partial pressure was kept at ~ $3.0 \times 10^{-6}$ Torr during growth. *In situ* reflection high-energy electron diffraction (RHEED) was used to monitor the overall growth rate and surface structure. After deposition, the substrate temperature was lowered at a rate of 50°C/min while the background $O_2$ was pumping out. The pure $SrCrO_3$ films were subjected to an additional anneal in air at 250°C for 3 hours to insure complete oxidation and formation of the perovskite phase.[22] This step was found to not be important for the LSCO films because upon air annealing, the changes in resistivity were negligible (see Fig. S1).[23]

High-resolution XPS using monochromatic Al K$\alpha_1$ x-ray ($h\nu$ = 1486.6 eV) was carried out at normal emission (electron take-off angle = 90° relative to the surface plane) with a VG/Scienta SES 3000 electron energy analyzer in an appended chamber. The total energy resolution was 0.5 eV. The binding energy scale was calibrated using a polycrystalline Au foil placed in direct



electrical contact with the film surface on the bench after deposition. XAS were measured in total electron yield (TEY) mode at beamline U4b of the National Synchrotron Light Source (NSLS) by measuring the sample drain current and were normalized to the current from a reference Au-coated mesh in the incident photon beam. The energy resolution was set at 180 and 260 meV for the O K and Cr $L_{3,2}$ edges, respectively. The photon energy scales of the O K- and Cr $L_{3,2}$-edges were calibrated using a method discussed elsewhere.[24] Hard x-ray XPS (HAXPES) at $h\nu$ = 4000 eV were measured at beamline X24a of the NSLS. The total energy resolution for these experiments was 0.5 eV. Microstructure and lattice parameters were determined using high-resolution x-ray diffraction (XRD) with a Philips X'Pert diffractometer equipped with a Cu anode. A hybrid monochromator, consisting of four-bounce double crystal Ge (220) and a Cu x-ray mirror, was placed in the incident beam path to generate monochromatic Cu $K\alpha$ x-rays ($\lambda$ = 1.54056 Å) with a beam divergence of 12 arc seconds. The electrical resistivity measurements were performed using van der Pauw method in the temperature range of 100-330 K. The cation compositions of films in the alloys series were determined by Rutherford backscattering spectrometry (RBS) using 2 MeV He ions.

All calculations were performed using the periodic model. The LSCO lattice was represented using a 2×2×2 cubic perovskite supercell. The Sr concentration was varied in increments of 12.5%. In each case, all non-equivalent geometrical arrangements of the La and Sr atoms as well as several spin-configurations within the Cr sublattice were considered. The calculations were performed using the Vienna *Ab initio* Simulation Package (VASP).[25,26] The projected augmented wave (PAW) method was used to approximate the electron-ion potential.[27] Exchange-correlation effects were treated within the Perdew−Burke−Ernzerhoff (PBE) functional form of the GGA, modified for solids (PBEsol).[28] The plane-wave basis with a 500



eV cutoff was used. Most of the calculations were performed using the 4 × 4 × 4 Monkhorst-Pack k-point mesh with its origin at the Γ point; the 8 × 8 × 8 mesh was used for selected cases. The charge and spin density distribution was analyzed using the Bader method.[29] The calculations were performed in two steps. First, the total energy was minimized with respect to the lattice parameters and internal coordinates. The energies of self-consistent calculations were converged to $10^{-6}$ eV/cell, and the lattice and atomic positions were relaxed until the forces on the ions were less than 0.02 eV/Å. It is well-known that the GGA does not describe localization of holes.[30] In our case, this shortcoming is manifested by delocalization of holes over all Cr species equally at all concentrations. To insulator-to-metal transitionigate this problem, we have used the GGA+$U$ approach to describe the on-site electronic correlations of the Cr 3$d$ orbitals, where the parameter $U$ was varied between 1 and 4 eV. The value $U = 3$ eV was found to result in the best agreement with the spectroscopic measurements.

**III. Results and discussion**

LCO exhibits an orthorhombic structure (space group *Pbnm*) with lattice parameters $a = 5.513$ Å, $b = 5.476$ Å, and $c = 7.759$ Å at room temperature. However, LCO can also be viewed as a pseudocubic cell with $a_{LCO} = 3.885$ Å. Bulk SCO adopts a cubic perovskite structure with a lattice constant of $a_{p\text{-}SCO} = 3.819$ Å.[22] Assuming a linear Vegard's law type dependence of lattice parameters on composition, we can estimate the lattice parameters for La$_{1-x}$Sr$_x$CrO$_3$ films with different x and their lattice mismatch with LAO ($a_{LAO} = 3.796$ Å) substrate. Doing so shows that the in-plane mismatch on LaAlO$_3$(001) should range linearly from +2.3% at x = 0 to +0.61% at x = 1. Assuming the critical thickness is not exceeded, the tetragonality and, thus, the strain should decrease as x increases.



RHEED patterns for the LSCO film series are shown in Fig. 1. The epitaxial nature of all films is clearly seen. No polycrystalline rings characteristic of disordered secondary phases are present. XRD direct-space maps along with out-of-plane θ-2θ scans (See Figs. S2 and S3[23]) were used to extract unit cell volumes for each film and these are plotted in Fig. 2, along with expectations based on two limiting cases – full relaxed bulk-like and fully coherently strained films. The larger error bars are due to the presence of a range of lattice parameters being present due to partial relaxation. With the exception of the x = 0.50 film, all exhibit unit cell (u.c.) volumes intermediate between the two limiting cases, indicating partial relaxation; the x=0.50 film is fully relaxed. None of the solid solution films exhibits an anomalously large u.c. volume as occurs when a high concentration of oxygen vacancies is present, resulting in chemical expansivity along the unclamped $c$ direction.[22]

We show in Fig. 3 core-level spectra for all four elements in LSCO for different x. The La $4d$ and Sr $3d$ line shapes are not affected by valence changes resulting from Sr doping whereas the Cr $2p$ line shape is strongly affected. The high-binding energy side of the O $1s$ spectra contain additional peaks associated with adsorbed $H_2O$, OH and organic contamination on the surface resulting from post-growth air exposure required to make the Au contact. Based on the relative intensities of the contaminant and lattice O $1s$ peaks, along with C $1s$ intensities, we estimate the thickness of the contamination layer for all samples except the x = 0.75 film to be of the order of 0.5 - 1 nm. For x = 0.75, the contamination thickness is ~ 2 nm. Moreover, this contamination is easily removed by room-temperature $O_2$ plasma treatment in the MBE chamber, establishing that it is surface bound, and not an integral part of the film. However, doing so over-oxidizes Cr in the near-surface region to $Cr^{6+}$, thereby obscuring vital information on the conversion of $Cr^{3+}$ to $Cr^{4+}$ with increasing x. Therefore, we have chosen to not take this cleaning



step. However, the low-binding energy side of the O 1$s$ lattice peaks is unaffected. Thus, the La 4$d$, Sr 3$d$ and O1$s$ spectra can be used to track changes in the chemical potential with x.[31] All three core levels shift to lower binding energy with increasing x. We use the inflection point on the low binding energy side of the O 1$s$ peak, rather than the peak position, because the overall line shape is skewed due to the appearance of OH and $H_2O$ peaks to higher binding energy.

The core-level binding energy shift ($\Delta E$) is given by $\Delta E = -\Delta\mu + K\Delta Q - \Delta V_M + \Delta E_R$, where $\Delta\mu$ is the change in chemical potential, $\Delta Q$ is the change in the number of valence electrons on the photoemitting atom, K is a constant, $\Delta V_M$ is the change in the Madelung potential, and $\Delta E_R$ is the change in the extra-atomic relaxation energy caused by the screening of the core-hole by metallic conduction electrons, or by polarization of surrounding media for insulators.[31] There is no change in the formal charge of O, Sr, and La as x increases, resulting in $\Delta Q = \sim 0$. The similar shifts seen in the O 1$s$, Sr 3$d$, and La 4$d$ peaks indicate that the change in the Madelung potential, which shifts anion and cation binding energies in opposite directions, is negligibly small. Core-hole screening by conduction electrons is also considered to be negligibly small in transition-metal oxides.[32] Therefore, the common shift measured for the O 1$s$, Sr 3$d$, and La 4$d$ core levels is largely due to the shift in chemical potential. We thus take the average of the three core-level shifts as a measure of $\Delta\mu$. Fig. 4a plots $\Delta\mu$ averaged over the three core levels as a function of x and indicates a monotonic downward shift in chemical potential with hole doping. This trend is similar to that seen in hole-doped early 3$d$ transition metal oxides such as $La_{1-x}Sr_xTiO_3$ and $La_{1-x}Sr_xMnO_3$.[31,33] The trend in chemical potential nicely parallels the trend in room-temperature resistivity, which is shown in Fig. 4b.

Returning to Fig. 3, the Cr 2$p$ spectra are rather complex by virtue of multiplet splitting and shake-up, in addition to the modified valence induced by Sr doping, changes in the chemical



potential with x, and the presence of $Cr^{6+}$ peaks at ~579.5 eV and ~588.5 eV for the films with x = 0.12, 0.50 and 0.85.[34] The latter result from surface oxidation of B-site Cr during the brief air exposure required to place the film surface in electrical contact with the grounded Au foil. This feature is not visible in the Cr 2p spectrum measured at $h\nu$ = 4000 eV (not shown) for which the electron attenuation length is ~3.3 times larger than it is at $h\nu$ = 1486.6 eV, indicating that the $Cr^{6+}$ is confined to the surface.[35] In order to extract information on the mix of $Cr^{3+}$ and $Cr^{4+}$ for the different x values, we thus focus primarily on line shape changes, rather than on absolute binding energies. The centroid of the Cr $2p_{3/2}$ peak for pure $LaCrO_3$ ($Cr^{3+}$) is at 576.0 eV, in good agreement with that for epitaxial $\alpha$-$Cr_2O_3$/Pt(111).[36] The intensity in the 576–578 eV range gradually increases with increasing x. This systematic change is due to an increase in the $Cr^{4+}$ concentration as more $Sr^{2+}$ substitutes for $La^{3+}$.[22] This trend is similar to that seen in the $Y_{1-x}Ca_xTiO_3$,[37] and $La_{1-x}Ca_xVO_3$,[7] in which the concentrations of $Ti^{4+}$ and $V^{4+}$ increase with $Ca^{2+}$ substitution as holes are doped onto 3d transition metal sites. In contrast, there are no striking changes in the Mn, Fe or Cu 2p line shapes with Sr doping in $La_{1-x}Sr_xMnO_3$,[33] $La_{1-x}Sr_xFeO_3$,[38] and $La_{2-x}Sr_xCuO_4$.[32] The latter are charge-transfer insulators according to the Zaamen-Sawatzky-Allen (ZSA) classification scheme and the top portion of the VB is primarily of O 2p character.[39] As a result, holes from Sr doping are injected primarily into O 2p-derived states, leaving the transition metal valences largely unchanged. The observation of a gradual increase in Cr valence from 3+ to 4+ with increasing x is consistent with the top of the VB be primarily of Cr 3d $t_{2g}$ character in LCO.[19] The observation of a $Cr^{4+}$ fraction in direct proportion to the Sr concentration does not necessarily mean that holes accompanying Sr doping are trapped on Cr sites. Itinerant holes on Cr sites in $La_{1-x}Sr_xCrO_3$ will also lead to the same fraction of Cr ions exhibiting Cr 2p spectra characteristic of $Cr^{4+}$ because the time scale core-level photoemission



well above threshold is much shorter than the time required for free carriers (either holes or electrons) to move from one lattice site to another, as has been observed for $Sr_{1-x}La_xTiO_3$.[40]

The change in Cr valence with Sr doping level and the associated effects on electronic structure are clearly seen in Cr L-edge and O K-edge XAS, as shown in Figs. 5. These spectra are not affected by changes in the chemical potential, and the higher energy resolution aids in observing the transition from $Cr^{3+}$ to $Cr^{4+}$. As x increases, the centers of gravity of the Cr $L_2$ and $L_3$ peaks move to higher x-ray energies. Moreover, the multiplet splitting pattern changes from one characteristic of $Cr^{3+}$ for LCO to one representative of $Cr^{4+}$ for SCO. Indeed, the latter matches well the Cr $L_{2,3}$ spectrum published for bulk, polycrystalline $CaCrO_3$.[41] The O K-edge spectra change dramatically as x increases. The Cr L-edge spectral line-shape for LCO ($Cr^{4+}$) and its evolution to SCO ($Cr^{3+}$) is consistent with earlier polycrystalline LSCO studies by Sarma *et al.*[16] We also observe the same set of O K-edge peaks, labeled A1-A3 ($LaCrO_3$) and B1-B3 ($La_{0.5}Sr_{0.5}CrO_3$) in Fig. 5, published earlier.[16] These spectra probe the transition from O 1$s$ to unoccupied states with at least partial O 2$p$ character hybridized with Cr 3$d$ states. The interaction of the core hole with the valence electrons leads to a perturbation to the electronic structure of the final state, complicating direct comparison of XAS spectra with the ground-state density of states. While this perturbation has been found to be substantial for transition metal L-edge XAS, detailed calculations of the effect of the O1$s$ core hole in transition metal oxides demonstrate that its effect on the band structure is weak.[42,43] Thus, the O K-edge spectra can be qualitatively related to unoccupied density of states of primarily transition metal character, provided there is enough hybridization with O 2$p$ to generate measurable oscillator strength. Indeed, new unoccupied states appear at lower x-ray energies as the VB is doped with holes accompanying Sr doping. Specifically, a new feature at ~529 eV becomes visible at the lowest



alloying level investigated (x = 0.12). As argued below, we assign this feature to a Cr 3*d*-derived band split-off from the VB proper in LCO. This new band results from hole doping, and it falls at a lower binding energy than the bottom of the conduction band proper, thus placing it in the gap.

The VB XPS and the O K-edge XAS data can be combined to provide a glimpse into the evolution of the electronic structure of both occupied and unoccupied bands as x increases. However, these spectra must first be put on a common binding energy scale, and this is done in Fig. 6a. All XPS spectra are referenced to the Fermi level. The O K-edge XAS yields the energy difference between the O 1*s* orbital and unoccupied bands above the Fermi level with a least partial O 2*p* character. In order to put XPS and XAS on a common energy scale, the energies of a common core orbital probed in both experiments (in this case, O 1*s*) must be known. The interaction of the excited electron and its core hole is in general greater for XAS than in XPS well above threshold by some amount, and we will call this quantity $\delta$. Thus, if $E_c$ is the energy of some empty conduction band state populated by the O K-shell x-ray absorption process (and negative by sign convention), and $h\nu_{XAS}$ is the corresponding x-ray energy based on a properly calibrated monochromator, then $h\nu_{XAS} + \delta + E_c = E_{XPS}^{O1s}$. We have used $\delta$ = 1.0 eV for LCO and all LSCO alloys, and $\delta$ = 0.7 eV for pure SCO. For the O K-edge, a rigid shift of 0.5 -1 eV has been employed for directly comparing the XAS with the DFT partial density of states for *p*-type oxides.[44] The value for SCO is smaller than that for the alloys and pure LCO because of more effective core-hole screening by conduction band electrons in SCO. Based on this analysis, we plot in Figs. 6a XPS VB and O K-edge XAS on a common energy scale and relabeled our absorption peaks to include the photoemission peaks (A-H).



Before discussing the effects of Sr doping, we first consider the spectra for pure LCO shown at the bottom of Figs. 6a. The VB spectrum for LCO consists of three features labeled A at ~ 6 eV, B at ~ 3.2 eV and C centered at ~ 1.5 eV. DFT calculations reveal that feature A is mostly O $2p$ derived with a minor Cr $3d$ contribution arising from O $2p$-Cr $3d$ hybridization. Feature B is O $2p$ non-bonding derived, and feature C is dominated by the occupied Cr $3d$ $t_{2g}$ orbital in a high-spin configuration with a minor admixture of O $2p$ character. The three spin-up (↑) Cr $3d$ $t_{2g}$ states are occupied and form the top of valence band, while other Cr $e_g$↑, $t_{2g}$ ↓ and $e_g$ ↓ states are unoccupied and form the bottom of conduction band. Therefore, we assign the peak labeled as D at -4 eV as due to excitation from O1$s$ to the unoccupied *Cr $e_g$↑ and $t_{2g}$ ↓* states hybridized with O $2p$. The broad feature at -6 eV (labeled E) is assigned to excitation to a hybridized La5$d$ derived state, with a small O2$p$-Cr $e_g$ ↓ component at lower energy. These assignments are consistent with our DFT band structure calculations and with assignments for LaCrO$_3$ powders.[16,45]

As can be seen by examining the complete O K-edge XAS data set in Fig. 5, the broad feature E (labeled B3 in Fig. 5) shifts to higher x-ray energy due to the change in orbital character from La 5$d$ to Sr 4$d$ with increasing x. Additionally, three new features (labeled as F, G and H in Fig. 6a) grow in intensity with increasing x and dominate the spectra for higher x values. It is useful to interpret these features by first analyzing the spectrum for SCO, where essentially all Cr is +4 and has a $(t_{2g}↑)^2$ electron configuration. The XAS spectral features for SCO correspond to excitation into the unoccupied $t_{2g}$↑, $e_g$↑, $t_{2g}$↓ *and* $e_g$ ↓ bands with a degeneracy ratio of 1:2:3:2. We note that the spectral intensity will be different from the degeneracy ratio because the XAS intensity depends on the extent of *p-d* hybridization. Thus, feature F in SCO is assigned to transitions to the unoccupied $t_{2g}$↑ band, G to a mixture of *$e_g$↑ and $t_{2g}$↓*, and H to $e_g$↓. These



assignments are in accordance with those of CaCrO$_3$[41] and CrO$_2$.[46] Both of these oxides contain Cr$^{4+}$ in octahedral coordination. We also note that our assignments are in agreement with those labeled A1, A2, A3, B1, B2, and B3 elsewhere [16] and highlighted in Fig. 5. Now, returning to LSCO, it is clear that feature F ($t_{2g}$↑) results from O K shell excitation to the unoccupied band introduced by Sr doping. This band grows in intensity with increasing x and moves to lower binding energy (e.g. away from the VB) with increasing x, remaining well above the Fermi level for all x. At the same time, features A, B and C in the VB spectra also shift to lower binding energies with increasing x. These shifts are consistent with the downward movement of chemical potential with hole doping, as deduced from the core-level spectra.

The correlated shifts of occupied band C and unoccupied band F to lower binding energies with increasing x is strongly suggestive of band F being a split-off component of band C resulting from electron depopulation of C as holes are added to the top of the VB. Sr doping in LCO thus converts LCO, an insulator, into a *p*-type semiconductor, as expected based on simple arguments. Measurable intensity at the Fermi level is observed for x ≥ 0.5, indicating degenerate *p*-type semiconducting, in accordance with the transport measurements.

The same evolution is seen in our PBEsol + $U$ calculations, which reveal that band F is a mixture of Cr 3$d$ $t_{2g}$ and O 2$p$. We show in Fig. 6b the theoretical densities of states (DOS) from PBEsol + $U$ with $U$ = 3 eV after convolving the occupied and unoccupied states with a Gaussian of full width at half maximum of 0.50 and 0.30 eV, respectively, to account for finite instrumental resolution in the XPS and XAS. As in the experimental data, the Cr 3$d$ $t_{2g}$ feature in the VB (C) moves toward the Fermi level as x increases and at the same time, a new feature appears above the Fermi level corresponding to the empty hybridized Cr 3$d$ $t_{2g}$ – O 2$p$ band resulting from hole doping. Removal of electrons from the Cr 3$d$ – O 2$p$ hybridized band at the



top of the VB (C) results in the generation of a virtual state in the gap (F), its energy being higher because it is unoccupied. As a result, low-lying optical excitation channels of nominal $d$-to-$d$ character are opened as Sr substitutes for La in the lattice, and these are clearly seen in the optical absorption spectra, which will be published in a separate paper on the optical properties of LSCO. The energy splitting between features C and F is best captured with $U = 3$ eV; a PBEsol calculation with $U = 0$ eV places these features much closer in energy than what we measure, indicating that correlation and the resulting localization of holes in important in describing LSCO.

Fig. 7 shows the temperature dependence of the electrical resistivity, $\rho(T)$, for the LSCO film series, along with fits of the data for x ≤ 0.50 to a model described below. Films with x ≤ 0.50 are clearly insulating or semiconducting (as defined using the criterion $d\rho/dT < 0$ for all $T$) with rapidly decreasing resistance at all temperatures as x increases. Undoped LaCrO$_3$ is not sufficiently conductive to be measureable using the conventional van der Pauw method ($\rho >$ ~900 Ω-cm) . The room-temperature (RT) resistivities for the doped films in the semiconducting regime ranges from 0.77 Ω-cm at x = 0.04 to 0.017 Ω-cm at x = 0.50 and all values are given in Table I. Seebeck measurements yield positive coefficients for x ≤ 0.50, indicating $p$-type conductivity. The RT resistivities of the epitaxial films are at least an order of magnitude lower than those for polycrystalline samples at comparable Sr doping levels except for x = 0, for which the MBE-grown LCO is much more resistive than the analogous powder samples.[13,14] Moreover, we find that unlike transport data for polycrystalline LSCO films taken from 300K to 2000K,[13,14] our data taken from 100K to 330K fit equally well to small polaron hopping ( $\rho \propto T\exp(\frac{E_a}{kT})$ ) and to band conduction ( $\rho \propto \exp(\frac{E_a}{kT})$) models. (See Fig. S4[23] and Table I). Additionally, the estimated activation energies deduced from these fits (Table I) are lower than those reported for



polycrystalline LSCO. These differences suggest that grain boundaries and electrically active defects and/or impurities play an influential role in determining transport properties in polycrystalline films. Epitaxial LSCO also exhibits an insulator-to-metal transition at x = 0.65 in the temperature range ~200-300 K over which $d\rho/dT > 0$, as also seen at x = 0.75, whereas polycrystalline LSCO exhibited no such transition, at least up to x = 0.50. The RT resistivities at higher x values of 0.65 and 0.75 are 0.003 and 0.002 Ω-cm, respectively.

Additionally, $d\rho/dT > 0$ for 50 K ≤ $T$ ≤ 300 K for pure, epitaxial SCO, indicating metallicity over a broader temperature range and a RT resistivity of ~0.001 Ω-cm. LCO is G-type antiferromagnet with a Neél temperature ($T_N$) of 290 K. It has been shown that $T_N$ decreases linearly with increasing Sr doping level, with $T_N$ = 200 K at x = 0.50.[17] In our resistivity data, the change of sign for $d\rho/dT$ occurs near this temperature for x = 0.65 and 0.75, but there is no discernible discontinuity in the region near $T_N$ for x ≤ 0.50, suggesting a weak coupling between the charge carriers and magnetism. The Sr-induced insulator-to-metal transition in LSCO resembles that observed in La$_{1-x}$Sr$_x$VO$_3$, although the critical Sr concentration necessary for the transition ($x_c$) is considerably higher in LSCO than in La$_{1-x}$Sr$_x$VO$_3$, for which $x_c$ = 0.17. Finally, we note that the insulator-to-metal transition is observed only for LSCO films deposited on LAO(001), for which the films are in compressive in-plane strain over the entire composition range; no insulator-to-metal transition occurs when an analogous film series is deposited on STO(001). In the latter, the LSCO films are in tensile strain for all x. The absence of strain effects may explain at least in part why polycrystalline bulk LSCO does not exhibit an insulator-to-metal transition.

Based on these results, we suggest that LSCO is best described in terms of *concurrent* band conduction and polaronic hopping, at least for temperatures up to ~300K. To investigate this



possibility, we fit the transport data for the semiconducting films (x ≤ 0.50) to a hybrid model in which we assume that polaron hopping involving trapped holes is frozen out below some temperature, whereas itinerant holes from band conduction are present over the entire temperature range. Specifically, we fit the transport data to the function

$$\rho(T) = \rho_0 e^{E_a/kT} + \frac{CTe^{E_a/kT}}{1+e^{[\frac{T_0-T}{\Delta T}]}} \qquad (1)$$

Here $T_0$ is the temperature at which polaron hopping freezes out, and $\Delta T$ is the temperature range over which this occurs. These fits are shown in Fig. 7 and are rather good over the entire temperature range and for all Sr concentrations. For all x values, $T_0$ is 100 – 120K. A single activation energy was used in both terms for simplicity, and its best-fit values are 0.100, 0.084, and 0.051 eV for x = 0.12, 0.25 and 0.50, respectively. In some sense, this model aligns well with early theoretical descriptions of polaron hopping. Holstein[47] considered the electron-phonon interaction required for polaronic motion as a small perturbation to the ground state of the system, which was taken to be a tight-binding-like electronic state in combination with a vibrational wavefunction. In this model, electrons at low temperatures diffuse to adjacent lattice sites within Bloch-like states with no change in the vibrational state of the system. This mode of transport is more band-like. As temperature increases, the band width drops due to the population of higher vibrational states and electron diffusivity diminishes. At approximately half the Debye temperature, the Bloch-like characteristics of the bands are lost and the mode of conduction changes from band-like to site hopping accompanied by changes in vibrational state, which is polaron hopping. This mode becomes completely dominant at sufficiently high temperatures and may explain why earlier high-temperature transport measurements for polycrystalline LSCO fit better to a polaron hopping model. However, there could be a competition between these two modes of transport at lower temperatures, as suggested by the present data set.

In order to understand polaron hopping in Sr-doped LCO more deeply, we used first principles modeling to correlate the local geometrical structure of Cr cations with the associated atomic charges in a $Sr_nLa_{8-n}Cr_8O_{24}$ supercell ($n$ = 0, 1,…,8). Here, the lattice parameters pre-determined using the PBE functional were fixed and only the internal coordinates were optimized using the PBE+$U$ approach with $U$ = 2.0, 3.0, and 4.0 eV. The latter two values of $U$ induce lattice distortions and a distribution of Cr atomic charges expected for polarons resulting from localized holes (see Figs. S5 and S6).[23] The shorter Cr-O bond lengths correspond to Cr cations having the larger ionic charge and a calculated moment of ~$2\mu_B$ per Cr, as expected for $Cr^{4+}$, whereas the longer Cr-O distances correspond to Cr species having the smaller ionic charge and a moment of ~$3\mu_B$, as expected for $Cr^{3+}$. In contrast, $U$ = 2.0 eV results in less charge localization and smaller lattice distortions, as evidenced by the hole charge and spin delocalization over several Cr sites (see Fig. S6).[23] A similar physical picture emerges if the PBEsol + $U$ functional is used instead.

As discussed above, the combined XPS and XAS measurements are best reproduced with the density of states obtained using $U$ = 3.0 eV (Fig. 6). We therefore conclude that a $U$ value of 3 eV is essential for accurately describing the electronic structure of LSCO, and thus that Sr-doping of LCO induces the formation of hole-based polarons. We note that at higher Sr concentrations, this electronic structure can be described just as well in terms of polarons resulting from localized electrons ($Cr^{3+}$) in a La-doped $SrCrO_3$ lattice.

To determine the energy barriers for $Cr^{3+} \rightarrow Cr^{4+}$ polaron hopping, we applied the linear interpolation method, in which the reaction coordinate connecting two energy minima ($\mathbf{R}_1$ and $\mathbf{R}_2$) is defined as $\mathbf{R} = \mathbf{R}_1(1 - t) + \mathbf{R}_2 t$, where $0 \leq t \leq 1$. This approach is appropriate for small atomic displacements. The energy barriers calculated using PBEsol+$U$ with $U = 3.0$ eV for three Sr concentrations are shown in inset to Fig. 7. The shortest polaron hopping path that does not involve spin flip can be either along [001] or [110] lattice vectors, depending on the most favorable spin configuration (which, in turn, changes with Sr concentration). In addition, since both hole-based and electron-based polarons are mutually repulsive if their respective concentrations exceed one per supercell, their hopping is correlated. Specifically, the calculated energy barriers for x = 0.25 and 0.50 correspond to the simultaneous hopping of two holes. Additionally, there are several non-equivalent configuration of holes and Sr dopants at x = 0.25, 0.50, resulting in a range of calculated energy barriers depicted by the connecting vertical lines at these concentrations in the inset. As the inset shows, the energy barrier for this hop is ~0.1 eV per hole, which is consistent with the experimental transport results.

**IV. Summary**

We demonstrate that substituting Sr for La on the A site in epitaxial LaCrO$_3$(001) deposited on LaAlO$_3$(001) increases the electrical conductivity as a result of hole doping of the valence band, leading to an insulator-to-metal transition at 65% doping level. XPS reveals that the Fermi level drops down toward the VB with increasing Sr concentration. These changes in electronic structure are manifested by the formation of a new band above the Fermi level seen in both experiment (XPS and XAS) and theory (PBE + $U$). This evolution in a natural consequence of adding holes to the Cr 3$d$ $t_{2g}$ derived top of the valence band which has the effect of moving this unoccupied, split-off subband to higher energy, and a value of $U = 3.0$ eV is required to accurately theoretically account for its energy relative to the top of the valence band. The character of the charge and spin distributions obtained using this value of $U$ points to the existence of the polaron contribution to electron transport and the calculated energy barriers for this mode of transport are consistent with the results of the experimental measurements.

**Acknowledgements**

This work was supported by the U.S. Department of Energy, Office of Science, Division of Materials Sciences and Engineering under Award #10122. The work was performed in the Environmental Molecular Sciences Laboratory, a national science user facility sponsored by the Department of Energy's Office of Biological and Environmental Research and located at Pacific Northwest National Laboratory. The computational work was supported in part by the PNNL Laboratory Directed Research and Development program. L. F. J. P. acknowledges support from the National Science Foundation under DMR 1409912. We thank Drs. Arena (U4b) and Woicik (X24a) for access and assistance at their end stations, and Drs. Kevin Rosso and Tim Droubay for helpful conversations concerning transport data. Use of the National Synchrotron Light Source Brookhaven National Laboratory was supported by the U.S. Department of Energy, Office of Science, Office of Basic Energy Sciences, under Contract No. DEAC02-98CH10886. Beamline X24a is supported by the National Institute of Standards and Technology.

**Table I**

| x in $Sr_xLa_{1-x}CrO_3$ | $\rho$ at RT ($\Omega$-cm) | Correlation coefficient for semiconductor model | Correlation coefficient for small polaron model | $E_a$ for band conduction model (eV) | $E_a$ for polaron hopping model (eV) |
|---|---|---|---|---|---|
| 0.12 | 0.44 | 0.998843 | 0.997976 | 0.084 | 0.100 |
| 0.25 | 0.088 | 0.999575 | 0.998794 | 0.069 | 0.084 |
| 0.50 | 0.018 | 0.993798 | 0.999433 | 0.035 | 0.051 |

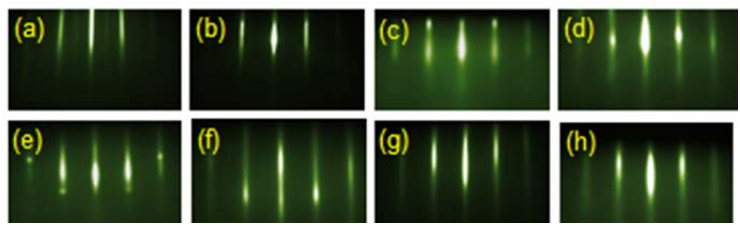

**Fig. 1** RHEED patterns for La$_{1-x}$Sr$_x$CrO$_3$/LaAlO$_3$(001) film set in the [100] zone axis: (a) x=0, 20 nm; (b) x=0.10, 80 nm; (c) x=0.25, 51 nm; (d) x=0.50, 62 nm; (e) x=0.66, 28 nm; (f) x=0.75, 54 m,; (g) x=0.85, 37 nm; (h) x=1.0, 42 nm.

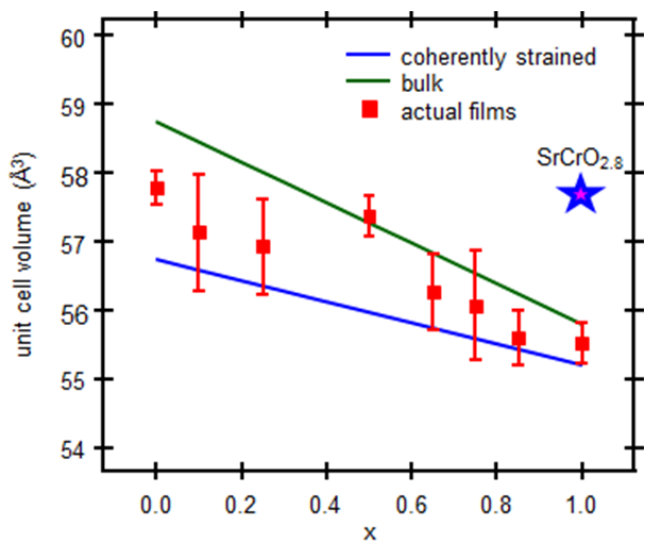

**Fig. 2** Unit cell volumes for the La$_{1-x}$Sr$_x$CrO$_3$/LaAlO$_3$(001) film set taken from XRD direct space maps (see Supplemental Document[23]) along with the expected behavior for two limiting cases: (i) films with bulk volumes assuming a linear relationship between the end members, and, (ii) fully coherently strained films using Poisson ratios for SCO and LCO and assuming a linear relationship. Also shown is a single data point measured for SrCO$_{2.8}$ deposited on LaAlO$_3$(001).

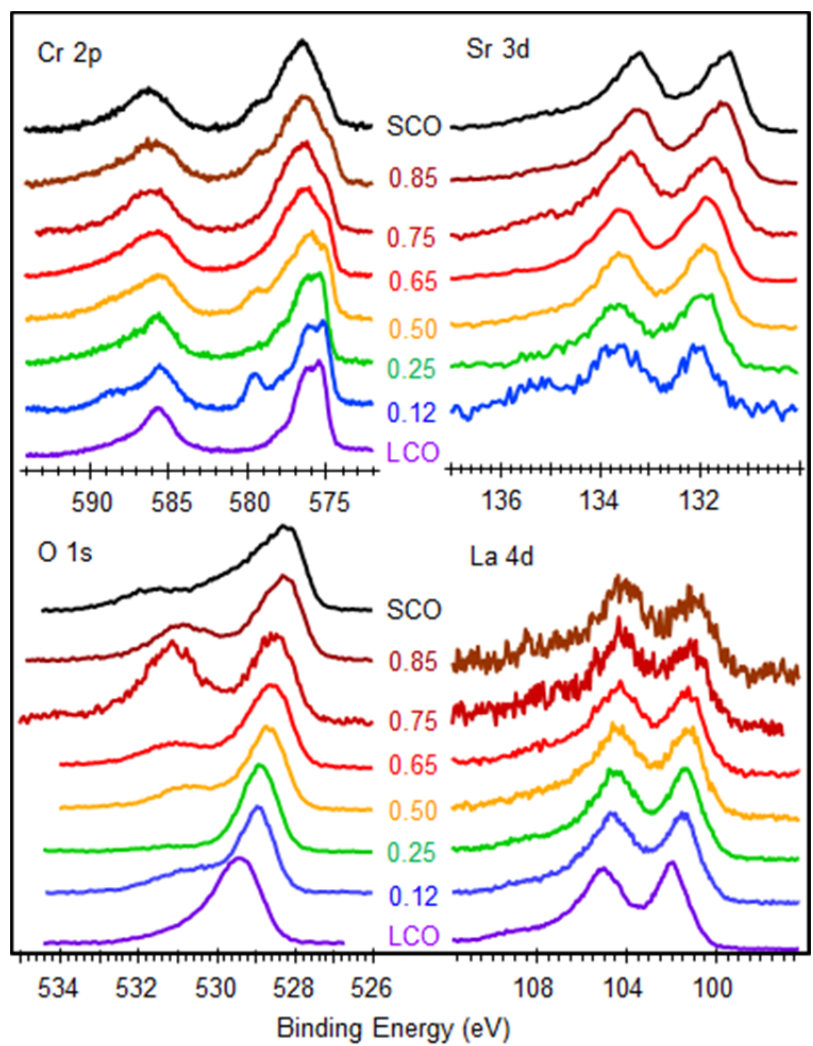

**Fig. 3** Cr 2*p*, O1*s*, La 4*d* and Sr 3*d* core-level spectra for $La_{1-x}Sr_xCrO_3$ as a function of x.

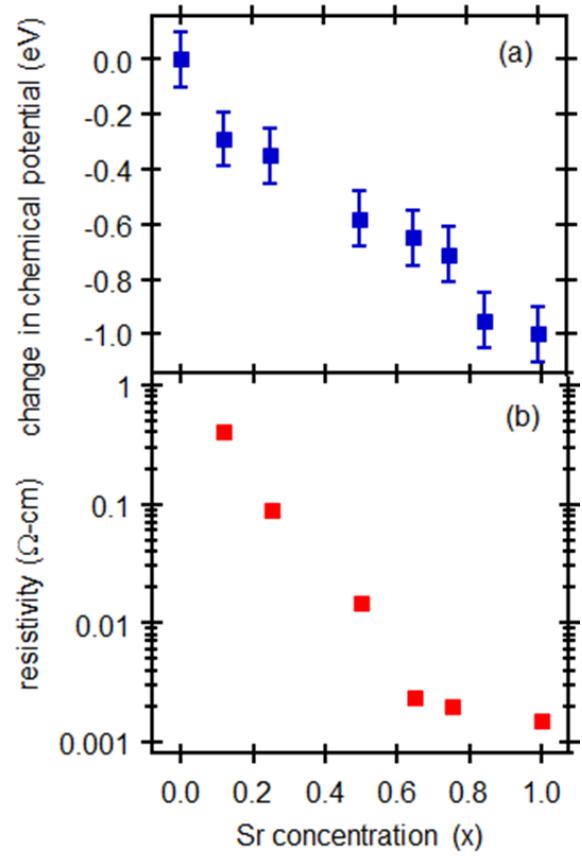

**Fig. 4** (a) Average chemical potential shift (Δμ) deduced from O 1$s$, La 4$d$ and Sr 3$d$ binding energy shifts and (b) room-temperature resistivity vs. x.

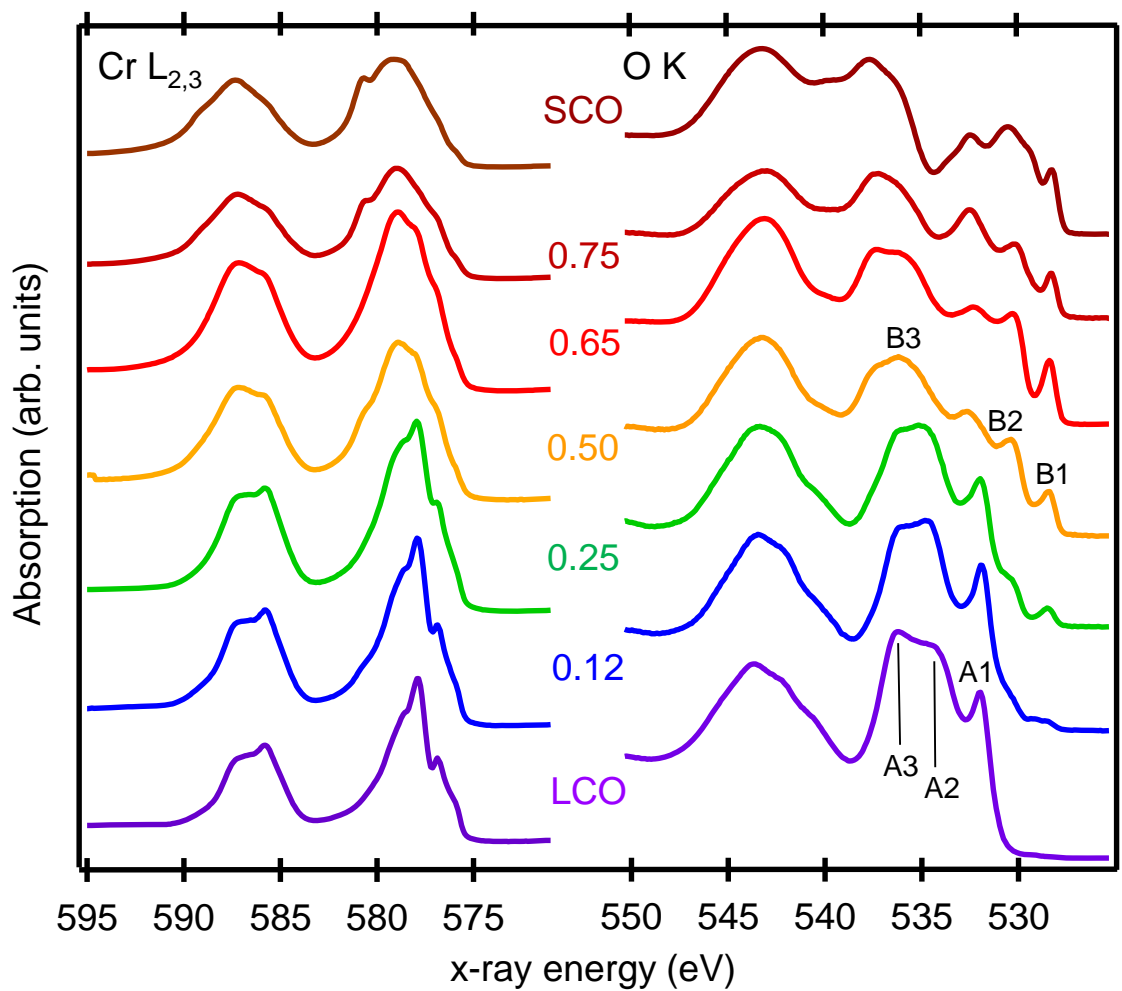

**Fig. 5** Cr L-edge and O K-edge XAS for the La$_{1-x}$Sr$_x$CrO$_3$ film series.

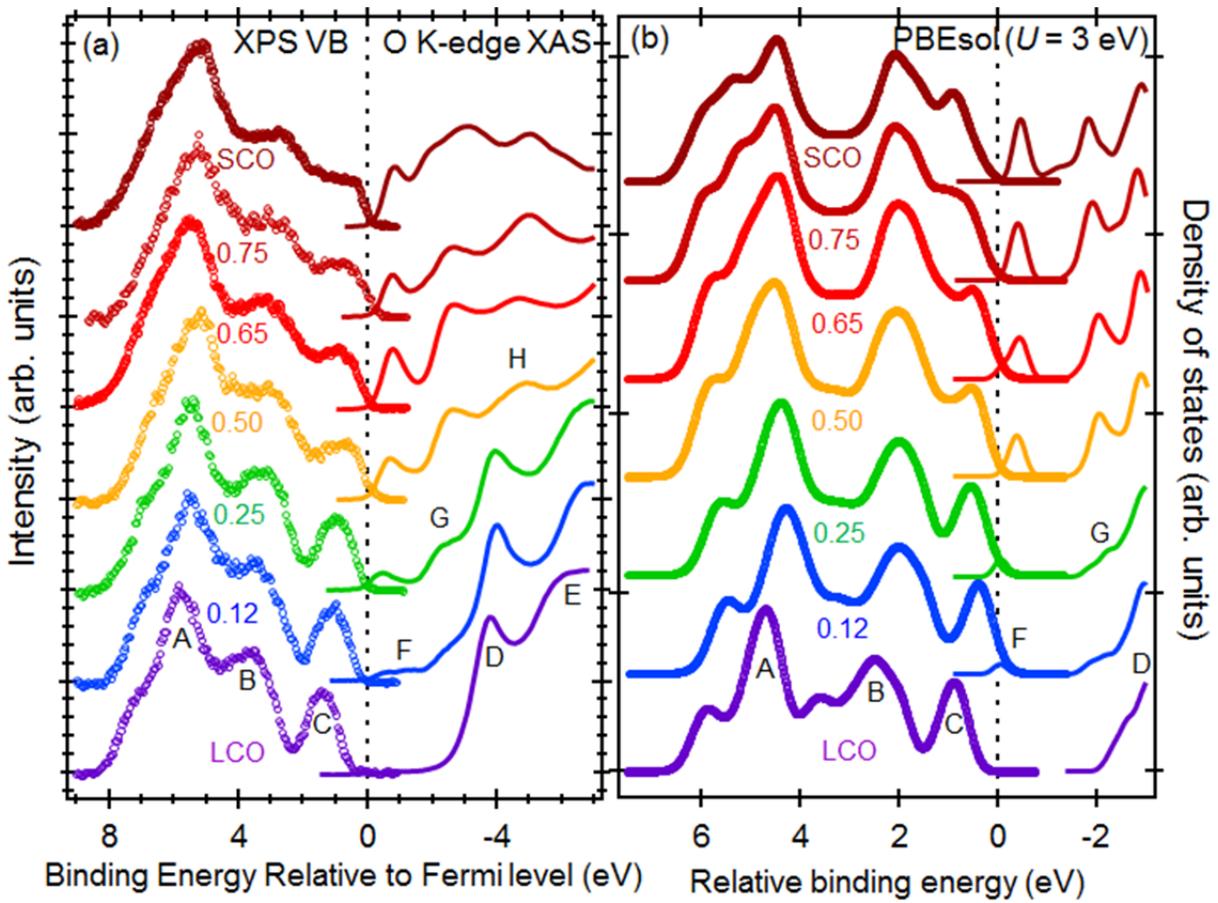

**Fig. 6** (a) Valence band XPS and O K edge XAS spectra for the La$_{1-x}$Sr$_x$CrO$_3$ film series; (b) analogous theoretical densities of states based on PBEsol + $U$ ($U$ = 3 eV).

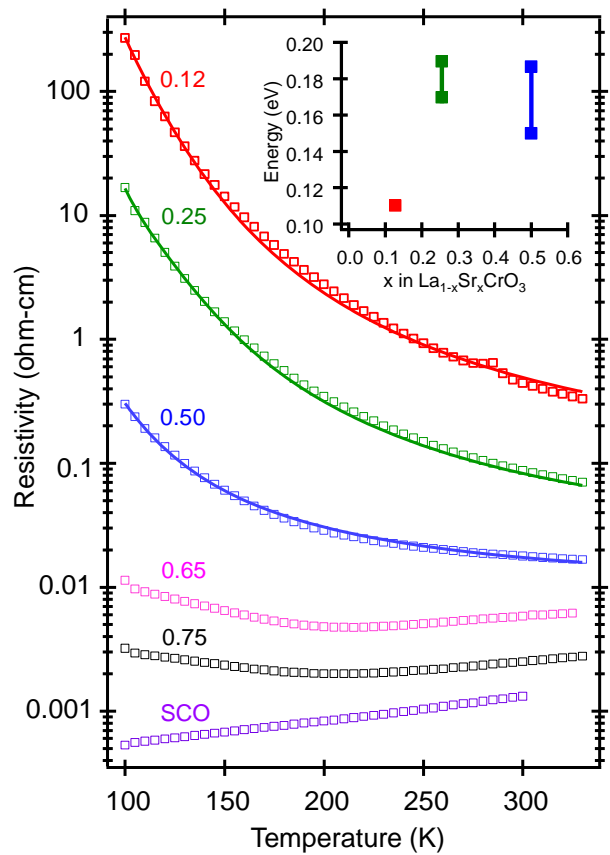

**Fig. 7** (a) $\rho(T)$ for the $La_{1-x}Sr_xCrO_3$ film series with the exception of pure $LaCrO_3$, which was too resistive to measure (open squares) along with fits of the data to eqn. 1 for x = 0.12, 0.25 and 0.50 (solid curves). Inset – calculated activation energies from a linear interpolation method based on PBEsol + $U$ (see text for details).

# Supplemental Document for

# Hole-induced insulator-to-metal transition in $La_{1-x}Sr_xCrO_3$ epitaxial films

K.H.L Zhang[1], Y. Du[2], P. V. Sushko[1], M. E. Bowden[2], V. Shutthanandan[2], S. Sallis[3], L.F.J. Piper[3], S.A. Chambers[1*]

[1]Physical Sciences Division, Fundamental and Computational Sciences Directorate, Pacific Northwest National Laboratory, Richland, WA 99352 U.S.A

[2]Environmental Molecular Sciences Laboratory, Pacific Northwest National Laboratory, Richland, WA 99352 U.S.A

[3]Materials Science & Engineering, Binghamton University, Binghamton, New York 13902, USA

*Corresponding author: sa.chambers@pnnl.gov

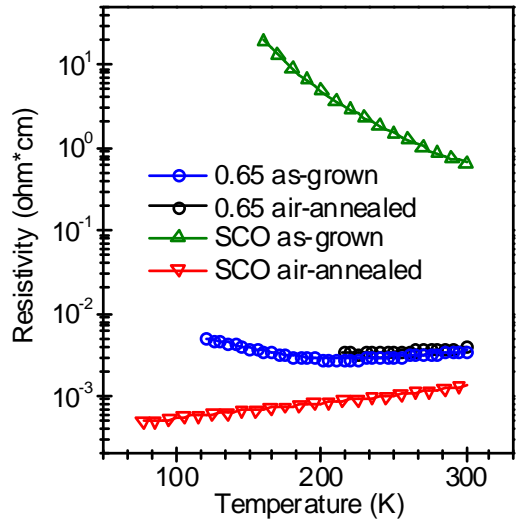

**Fig. S1** Effect of air annealing at 300°C for 2 hours on the resistivities of $La_{0.35}Sr_{0.35}CrO_3$(001) and SCO (001) epitaxial films. The resistivity of SCO decreases by several orders of magnitude and the unit cell volume decreases significantly (Fig. 2) as a result of annealing in air. This behavior is indicative of the healing of O vacancies as the material transitions from semiconducting, rhombohedral $SrCrO_{2.8}$ to metallic perovskite $SrCrO_3$. In contrast, neither the resistivity nor the unit cell volume change when $La_{0.35}Sr_{0.35}CrO_3$ is annealed in air, indicating that $La_{0.35}Sr_{0.35}CrO_3$ does not exhibit a high degree of O vacancies as grown.

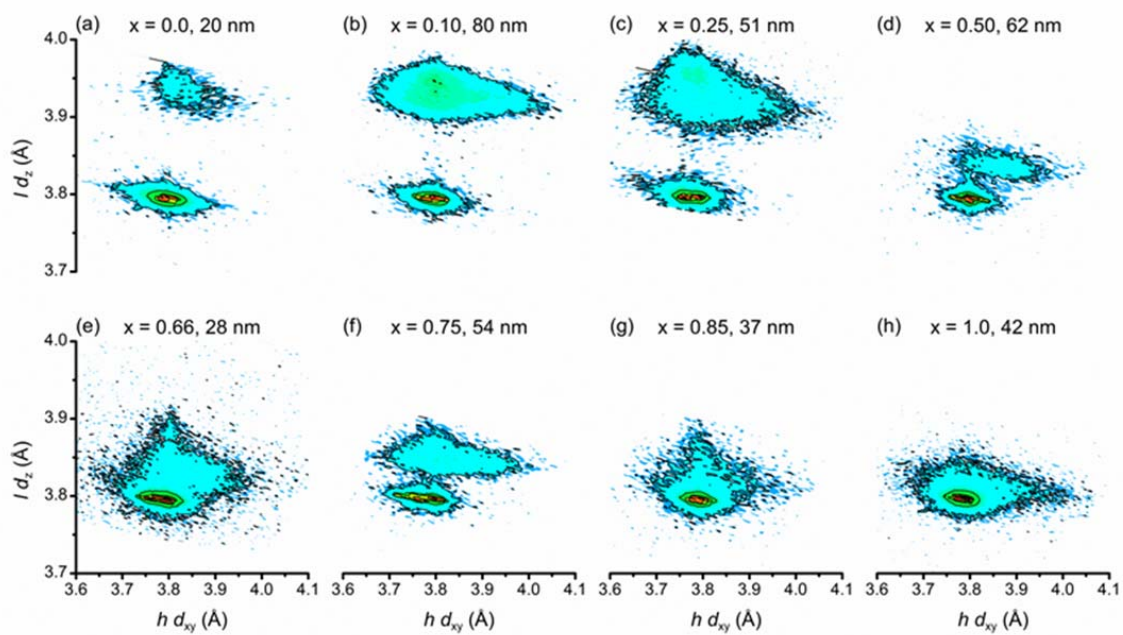

**Fig. S2** Direct space maps derived from XRD data for the $La_{1-x}Sr_xCrO_3$ alloy film series from which the lattice parameters in Fig. 2 are extracted. The film thickness is given for each composition (x). In some cases, the most noteworthy being (b) and (c), the films exhibit a range of lattice parameters for both in-plane and out-of-plane directions, characteristic of partial relaxation.

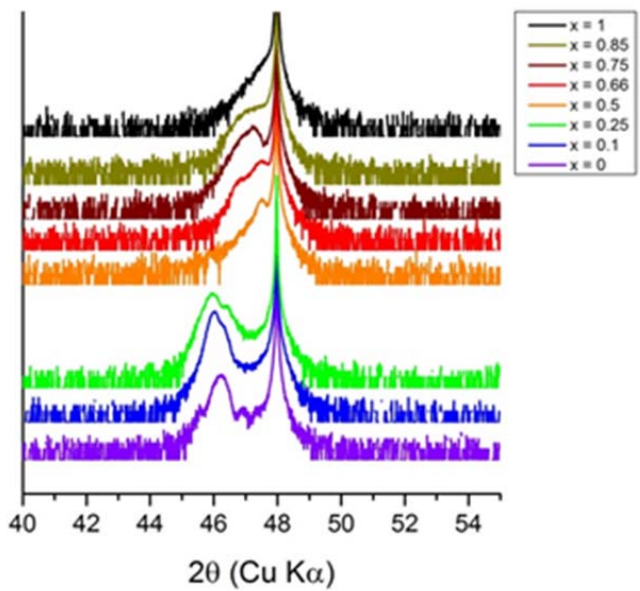

**Fig. S3** Out-of-plane $\theta - 2\theta$ XRD scans for the $La_{1-x}Sr_xCrO_3$ (001) alloy film series from which the lattice parameters in Fig. 2 are extracted.

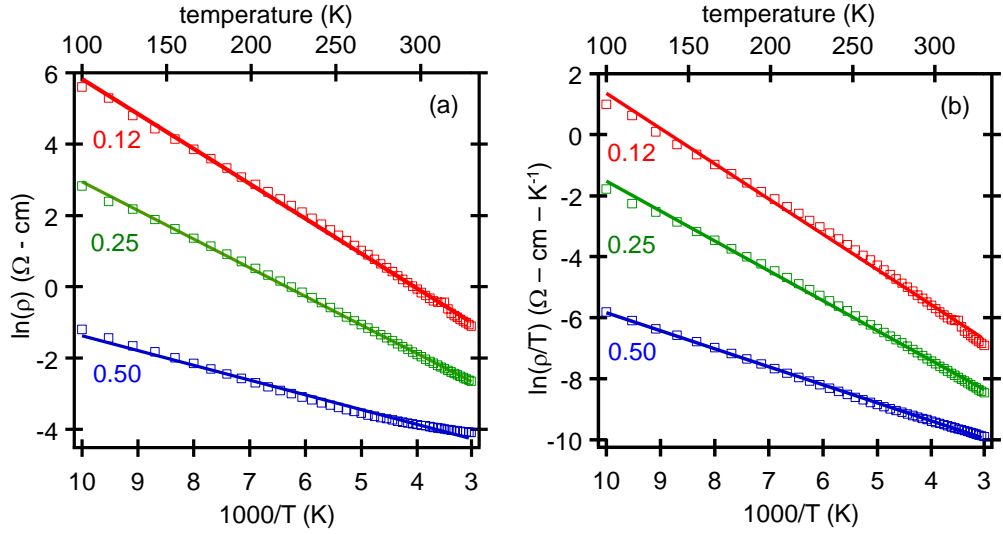

**Fig. S4** Fits of resistivity data to band conduction (a) and polaronic hoping (b) models for LSCO films with x ≤ 0.5.

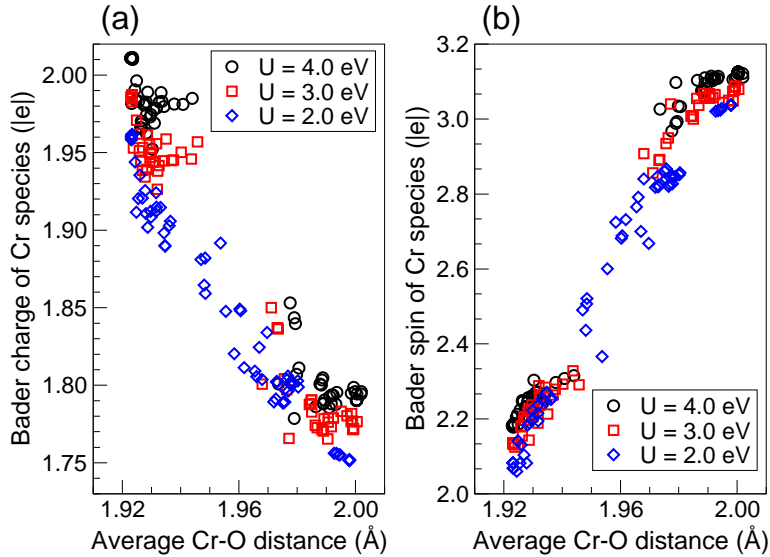

**Fig. S5** Correlation between the average Cr-O distance for each Cr species and the associated atomic charge (a) and spin (b) in $Sr_nLa_{8-n}Cr_8O_{24}$ ($n$ = 0,1,2,…,8). The charge and spin per Cr atom were calculated using the Bader population analysis. For $U$ = 3.0 and 4.0 eV, the average Cr-O distances larger than 1.96 Å correspond to a calculated Cr Bader charge of 1.8 |e| and magnetic moment of ~$3\mu_B$, which corresponds to formally $Cr^{3+}$ ions. Conversely, if the Cr-O distances are smaller than 1.96 Å, the Bader charges of the Cr species increase to up to 2.0 |e| and the spins decrease to ~$2\mu_B$, which correspond to $Cr^{4+}$. Thus, typical of polarons, localization of the holes on Cr cations is accompanied by strong lattice deformation. For $U$ = 2.0 eV, the hole localization is less pronounced, as manifested by the continuous values of the average Cr-O distances and the corresponding Cr charges.

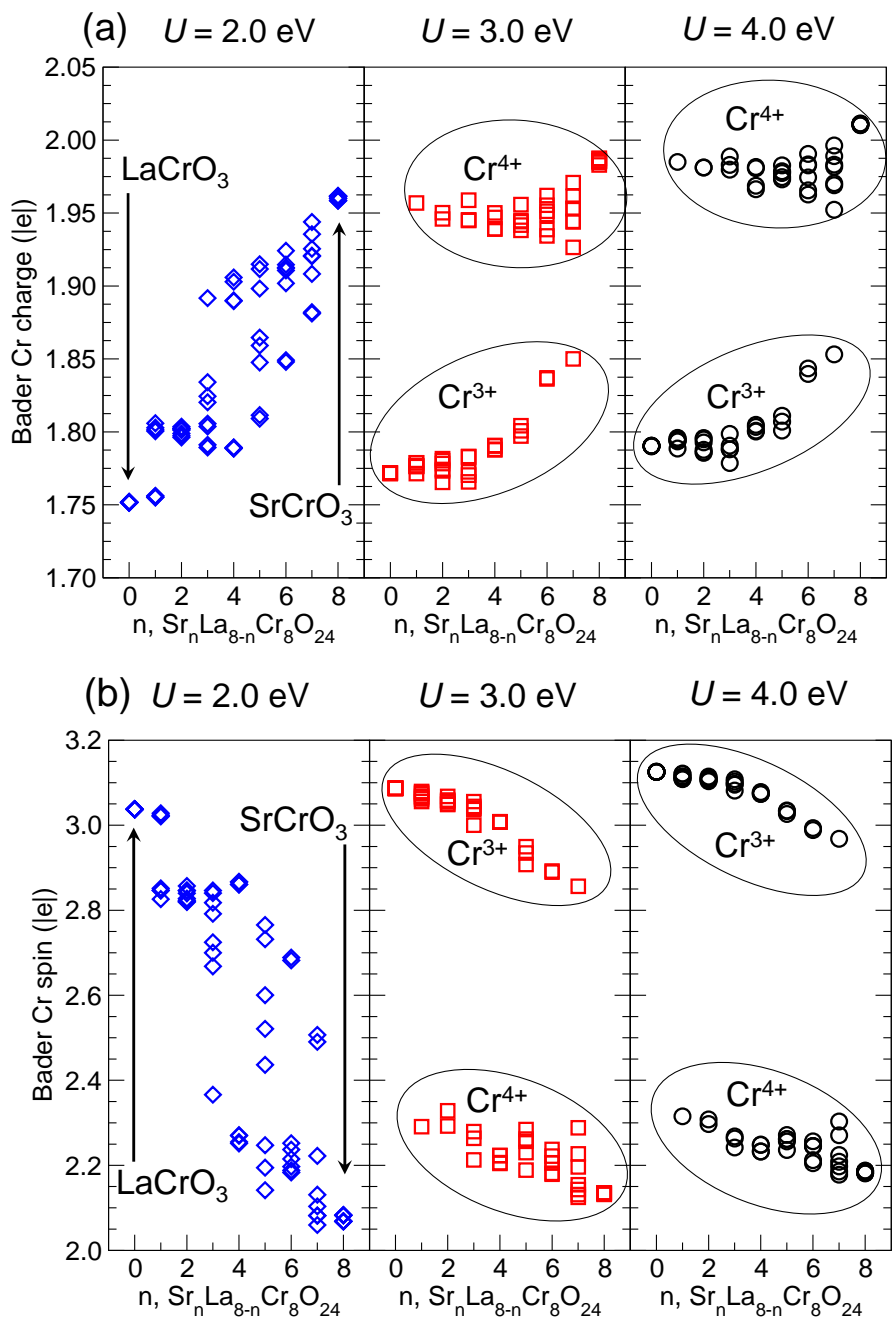

**Fig. S6** Charge (a) and spin (b) of Cr cations as a function of the Sr concentration and the value of the Hubbard $U$ correction. The charge and spin per Cr atom were calculated using the Bader population analysis.

4